\documentclass[arXiv,onecolumn,
superscriptaddress,showpacs]{revtex4}
\usepackage[intlimits]{amsmath}
\usepackage{amsfonts}
\usepackage{graphics}
\usepackage{subfigure}
\usepackage[usenames]{color}
\usepackage{epstopdf,epsfig}

\usepackage{indentfirst}

\usepackage{graphicx}
\usepackage{amsmath}
\usepackage{amsfonts}
\newcommand\be            {\begin{equation}}
\newcommand\bea           {\begin{equation}\begin{array}l\displaystyle}
\newcommand\ee            {\end{equation}}
\newcommand\bes           {\begin{subequations}}
\newcommand\esu           {\end{subequations}}

\newcommand\p            {\partial}

\def\3pt#1#2#3{{\langle{#1}\vert{#2}\vert{#3}\rangle}}

\begin{document}
\title{
Deviations from Off-Diagonal 
Long-Range Order in One-Dimensional Quantum Systems}

\author{A. Colcelli}
\affiliation{SISSA and INFN, Sezione di Trieste, via Bonomea 265, I-34136
Trieste, Italy}

\author{G. Mussardo}
\affiliation{SISSA and INFN, Sezione di Trieste, via Bonomea 265, I-34136
Trieste, Italy}
\affiliation{International Centre for Theoretical Physics (ICTP), 
strada Costiera 11, I-34151, Trieste, Italy}
\affiliation{Institute for Theoretical Physics, Utrecht University}

\author{A. Trombettoni}
\affiliation{CNR-IOM DEMOCRITOS Simulation Center, via Bonomea 265, I-34136 Trieste, Italy}
\affiliation{SISSA and INFN, Sezione di Trieste, via Bonomea 265, I-34136
Trieste, Italy}


\begin{abstract}
A quantum system exhibits off-diagonal long-range order (ODLRO) 
when the largest eigenvalue $\lambda_0$ of the one-body-density matrix 
scales as $\lambda_0 \sim N$, where $N$ is the total 
number of particles. 
Putting $\lambda_0 \sim N^{{\cal C}}$ to define the scaling exponent 
${\cal C}$, then ${\cal C}=1$ corresponds to 
ODLRO and ${\cal C}=0$ to the single-particle occupation of the 
density matrix orbitals. When $0<{\cal C}<1$, ${\cal C}$ can be used 
to quantify deviations from ODLRO. 
In this paper we study the exponent ${\cal C}$ in a variety of one-dimensional 
bosonic and anyonic quantum systems. For the $1D$ Lieb-Liniger Bose gas we find 
that for small interactions ${\cal C}$ is close to $1$, implying 
a mesoscopic condensation, {\em i.e.} a value of the ``condensate'' 
fraction $\lambda_0/N$ appreciable 
at finite values of $N$ (as the ones in experiments with $1D$ ultracold atoms). $1D$ anyons provide the possibility to fully interpolate between 
${\cal C}=1$ and $0$. The behaviour of ${\cal C}$ for these systems 
is found to be non-monotonic both with respect to  
the coupling constant and the statistical parameter.
\end{abstract}

\maketitle

\section{Introduction} The Penrose--Onsager criterion for the presence of 
off--diagonal 
long--range order (ODLRO) is the cornerstone of the present understanding 
of quantum coherence and Bose--Einstein condensation (BEC) 
\cite{Penrose56}. It is simply related to 
the occurrence of BEC 
and it is based on 
the study of the scaling with the number of particles of the eigenvalues of the one--body density matrix (1BDM) 
$\rho(x,y)$, defined as \cite{Pitaevskii16}:
\be
\rho(x,y)=\langle \hat{\Psi}^\dag(x) \hat{\Psi}(y) \rangle \, , 
\label{def}
\ee
where $\hat{\Psi}^\dag(x)$ 
is the field operator creating 
a particle at the 
point $x$. Denoting by $\lambda_i$ the eigenvalues of this matrix, we have
\be
\int dy \, \rho(x,y) \, \phi_i(y) = \lambda_i \, \phi_i(x) \, ,
\label{eig}
\ee
where the $\phi_i$'s are the corresponding eigenfunctions. One has ODLRO 
and BEC when the largest eigenvalue $\lambda_0$ scales as the total number of particles $N$ of the system. The occurrence of ODLRO implies phase coherence, as shown by a simple argument due to Anderson \cite{Anderson66} and reviewed in \cite{Huang95}. 
\\\indent The Penrose--Onsager criterion relates, altogether, the occurrence of BEC 
and quantum coherence to the behaviour of correlation functions. Its power and elegance stem from the fact that it applies at zero and finite 
temperatures and as well in any dimensions, so that in Eqs.\,(\ref{def})-(\ref{eig}), the coordinates $x,y$ may denote space vectors with $D$ components, possibly also including spin degrees of freedom. Moreover, the system may also be subjected to a generic one--body external potential. A major example 
of detection of ODLRO is provided by the measurement of the momentum 
distribution $n(k)$ 
in ultracold atom experiments, with a clear peak around zero momentum forming 
at the BEC critical temperature \cite{Anderson95}.
\\
\\\indent When the system is homogeneous and the thermodynamic limit is taken in the usual way by keeping fixed the density $n \equiv N/\Omega$ ($\Omega$ is the volume), then $\rho(x,y)$ tends to the condensate density $\lambda_0/\Omega$ when $|x-y| \rightarrow \infty$ \cite{Pitaevskii16}. This definition makes transparent the analogy of the condensate fraction $\lambda_0 / N$ with the magnetization $M$ in magnetic spin systems, where the analog of the 1BDM (\ref{def}) is the correlation function $\langle S_i S_j \rangle$ which, in the homogeneous case, tends to $M^2$ for $|i-j| \to \infty$ (see, for instance, \cite{Mussardo10}).
\\\indent Given that in presence of ODLRO 
the largest eigenvalue $\lambda_0$ scales as $N$, 
we can conveniently quantify deviations from ODLRO in terms of the 
exponent ${\cal C}$ of a scaling law as 
\be
\lambda_0 \sim N^{{\cal C}} \, .
\label{scal}
\ee
Clearly, when ${\cal C}=1$ we are back to the ODLRO and BEC, according to the Penrose-Onsager criterion. On the other hand, when ${\cal C}=0$, we are typically in a situation which is fermionic-like: think, for instance, at the ideal Fermi gas, where for {\em all} eigenvalues (including $\lambda_0$) we have $\lambda_i =1$, in view of the Pauli principle. As additional example, 
consider a system made of two species of fermions with attractive interactions 
where there may be ODLRO but this manifests in the two-body density matrix, 
while for the scaling law of the eigenvalues of the 1BDM one still has 
${\cal C}=0$. Despite one can imagine more general, non-power-law, 
dependence of $\lambda_0$ on $N$, it is reasonable to assume a power-law form 
like the one introduced in Eq.\,(\ref{scal}). The explicit 
computations presented below on one-dimensional systems are in agreement 
with the definition (\ref{scal}).
\\\indent One-dimensional quantum systems provide an ideal playground to investigate deviations from ODLRO since  there is no BEC in the interacting case. In other words, one expects ${\cal C}=1$ only for the $1D$ noninteracting Bose gas, which may be regarded, however, as a very delicate, if not pathological, limit. 
In the homogeneous case, in fact, any 
infinitesimal repulsive interaction, no matter how small, in $1D$ 
destroys ODLRO also at $T=0$, unlike the $3D$ case. This means that, 
in $1D$ interacting systems, ${\cal C}$ must be strictly smaller than 
$1$ for any finite value of the interaction. Therefore one may lead to 
conclude that no clear peak of the momentum distribution 
should be observed, also at $T=0$, in experiments 
with ultracold atoms in one-dimensional confined geometries. 
\\\indent However, when ${\cal C}$ is close to $1$, for finite $N$ the ratio 
$\lambda_0/N$ can be rather large 
(even though $\lambda_0/N$ tends to $0$ for $N \to \infty$). 
When this happens, we say that we are in presence of a {\em mesoscopic condensation}, {\em i.e.} a phenomenon that, for all practical purposes, can be considered as an ordinary condensation: {\em e.g.}, for ${\cal C} \approx 0.99$ and $N \approx 10^3$ one has $N^{{\cal C}}/N \approx 0.9$. This implies 
that even in absence of BEC, if ${\cal C}$ is rather close to $1$, one would 
observe a clear peak in the momentum distribution, 
especially because typically in experiments with $1D$ ultracold gases the 
number of particle is $N \sim 10^2-10^3$. Given 
the fact that the momentum distribution is an experimental 
quantity easily accessible, the study of deviations from ODLRO for different 
geometries and interactions is therefore desirable.
\\\indent The topic of this paper is to identify and quantify deviations from 
ODLRO in $1D$ quantum systems at $T=0$. Apart from the obvious consequences 
for the presence of a mesoscopic peak in the momentum distribution, 
there are three additional reasons for such a study. 
First of all, the computation of correlation functions and 
1BDM is a quite difficult and often formidable task. For $1D$ systems, 
however, the situation is generally  
better and a huge variety of techniques has been developed for this aim 
\cite{GiamarchiBook,Cazalilla11}, 
ranging from bosonization \cite{Haldane81} and density matrix renormalization group \cite{Schollwock05}, 
to Bethe ansatz and integrability techniques \cite{korepin,Franchini}. 
Secondly, one--dimensional anyonic gases set a non-trivial interpolation 
between Bose and Fermi statistics, and have the further advantage 
to be Bethe solvable \cite{Kundu99,Batchelor06}. 
Finally, ultracold atoms provide 
an ideal setting to simulate different $1D$ quantum systems 
by acting on tunable external parameters \cite{Yurovsky08,Cazalilla11}. 
For instance, the coupling constant $\gamma$ in $1D$ ultracold bosonic gases 
can be adjusted by tuning the transverse confinement of the waveguides in which the atoms are trapped \cite{olshanii98}, and in such an experiment one can explore both the regimes of small $\gamma$, as small as $10^{-4}-10^{-3}$ (the weakly interacting limit), and large $\gamma$ (the Tonks-Girardeau limit 
\cite{girardeau60}), 
with numbers of particles $N$ going from few tens to thousands, see the reviews \cite{Yurovsky08,Cazalilla11,Bouchoule09}. Let's remark that ${\cal C}$ being close to $1$ at small $\gamma$ 
gives reason to the fact that in the weakly interacting limit the mean-field description works reasonably 
well, despite the absence of a proper BEC. 

\section{$1D$ interacting Bose gas} Since fermions have always ${\cal C}=0$ due to their statistics, we start our analysis from the 
Lieb-Liniger (LL) model \cite{LL}, a homogeneous $1D$ system of 
$N$ bosons of mass $m$ interacting via a two-body repulsive 
$\delta$--potential in a ring of circumference $L$. The Hamiltonian reads
\be
\label{HAM_LL}
H_{LL}=-\frac{\hbar^2}{2m}\sum_{i=1}^N\frac{\p^2}{\p x_i^2}+ 2 c \,
\sum_{i<j}\delta(x_i-x_j) \, ,
\ee
and one defines a dimensionless coupling constant
\begin{equation}
\gamma=\frac{2\,m\, c}{\hbar^2 \, n} \, , 
\label{eq:gamma}
\end{equation}
where $n = N/L$ is the density of the gas, kept constant in the thermodynamic 
limit. 
As well known, the LL model is exactly solvable by Bethe ansatz \cite{LL,yang} which provides the exact expression of the many--body eigenfunctions \cite{korepin,Gaudin}. At $T=0$ the ground--state energy, 
the sound velocity $s$ and other equilibrium quantities, appropriately scaled, can be expressed in terms of the solution of the so-called Lieb integral equations \cite{LL}, which in turn depends only on $\gamma$. 
The equation of state coincides with the one of an ideal gas in the two limits $\gamma \to 0$ and $\gamma \to \infty$, with the residual energy depending in general on $\gamma$ \cite{Mancarella14}. The LL $1D$ Bose gas can be treated as well by bosonization. 
The dimensionless parameter called the Luttinger parameter $K$ 
\cite{GiamarchiBook} 
can be written for the LL model 
as $K =v_F / s$ 
where $v_F= \hbar \pi n/m$ is the Fermi velocity. 
Therefore, solving the Lieb integral equations one has access both to the sound velocity $s$ and the Luttinger parameter $K$ for any values of the coupling constant $\gamma$ (see, {\em e.g.}, \cite{CazalillaCitroGiamarchi} and 
\cite{Lang17} and Refs. therein). In particular in the weak coupling limit $\gamma \ll 1$ one has $s \simeq \frac{\hbar}{m} n \sqrt{\gamma}$ and $K \simeq \frac{\pi}{\sqrt{\gamma}}$. For the Tonks-Girardeau gas one has at variance $s = v_F$ and $K=1$, so that $K$ for a homogeneous system of $\delta$--repulsive bosons goes from $\infty$ (for $\gamma \rightarrow 0$) to $1$ (for $\gamma \rightarrow \infty$). 
\\\indent A simple evaluation of ${\cal C}$ using bosonization can be done as follows. The eigenvalues $\lambda_i$ and the orbitals $\phi_i(x)$ in Eq.\,(\ref{eig}), for (\ref{HAM_LL}), are labelled by a quantum number which is evidently the momentum $k$. From translational invariance $\phi_k(x) = (1/\sqrt{L}) e^{ikx}$ \cite{Pitaevskii16}, and therefore $\lambda_k$ simply equals 
the momentum distribution $n(k)=\langle \hat{\Psi}^\dag(k) 
\hat{\Psi}(k) \rangle$, where the operator $\hat{\Psi}(k)$ is the Fourier transform of the field operator $\hat{\Psi}(x)$ \cite{Pitaevskii16}. It follows $n(k) \propto \int_0^L dx\, \rho(x) e^{ikx}$ where, using translational invariance, we have set $y=0$ and $\rho(x,0) \equiv \rho(x)$. From Luttinger liquid theory for large $x$ we have that 
$\rho(x)\,\propto\,x^{-1/2K}$ and therefore $n(k) \propto \frac{1}{k^{1-1/2K}}$ for $k \to 0$ \cite{note1}. Since the 
smallest momentum is $k_{min}\propto2\pi/L$ and $n=N/L$, one gets $\lambda_0 \propto N^{1-1/2K}$, {\em i.e.} 
\be
{\cal C}=1-\frac{1}{2K} \, .
\label{res}
\ee
Notice that ${\cal C}$ is then expected to depend on $\gamma$, {\em i.e.} on the ratio $c/n$, and not on $c$ and $n$ separately.
\\\indent An accurate, high-precision check of such a prediction is not easy to obtain, 
since one should determine $\lambda_0$ as a function of $N$ 
and then fit ${\cal C}$ from the scaling. It is clear that 
the larger is the maximum value of $N$ considered, the better the estimate 
of ${\cal C}$, but from exact computations is not straightforward to reach 
large values of $N$, especially for intermediate values of $\gamma$. 
Of course, one could think to use the Bethe ansatz expression for the wave function of the ground state, but, in practice, even for small number of particles, such an expression is difficult to handle. One way to get around this difficulty consists, {\em e.g.}, in using a numerical approach as ABACUS \cite{CauxCalabrese2006, PanfilCaux2014}, in which the sum on the corresponding Bethe eigenfunctions can be efficiently truncated. In \cite{CalabreseABACUS} such a method was used for the approximate computation of the 1BDM up to $150$ particles and for any values of $\gamma$. In the following we introduce a new method based on an interpolation of the 1BDM between large and short distance asymptotic expansions, and which can be used to larger number of particles for all values of $\gamma$. Hereafter we present the results obtained with the interpolated $\rho(x)$ till $N=10^3$ (there is however no major problem to extend such a computation to larger values of $N$). For $N=10^{3}$ we found that the error in ${\cal C}$ is, 
{\em e.g.}, at the fifth significant figure for $\gamma=1$ and such an error can be further decreased since larger the value of $N$, 
smaller the error in ${\cal C}$. Our results with $N=10^3$ 
confirm that the ${\cal C}$ 
depends only on $\gamma$ and not separately on $n$ and $c$. 
\\\indent To conveniently set up such an interpolation formula we have built upon several known behaviors of the 1BDM $\rho(x)$ as a function of distance $x$, and in particular on its short-distance  
behaviour $\rho^{SD}(x)$ \cite{Olshanii-short} and on the large-distance 
one, $\rho^{LD}(x)$ \cite{Shashi2011,PanfilCaux-large}. 
The limits of weak and strong interactions, 
valid for any values of $x$, have been extensively investigated 
\cite{JimboMiwa,CastinMora,Gangardt2003,Forrester06,Imambekov2009}
(see more Refs. in 
\cite{Cazalilla11,Yurovsky08}). Such 
known expressions are collected for convenience in the Appendix. 

\section{Interpolation scheme} 
\begin{figure}[t]
\centerline{\scalebox{1.0}{\includegraphics{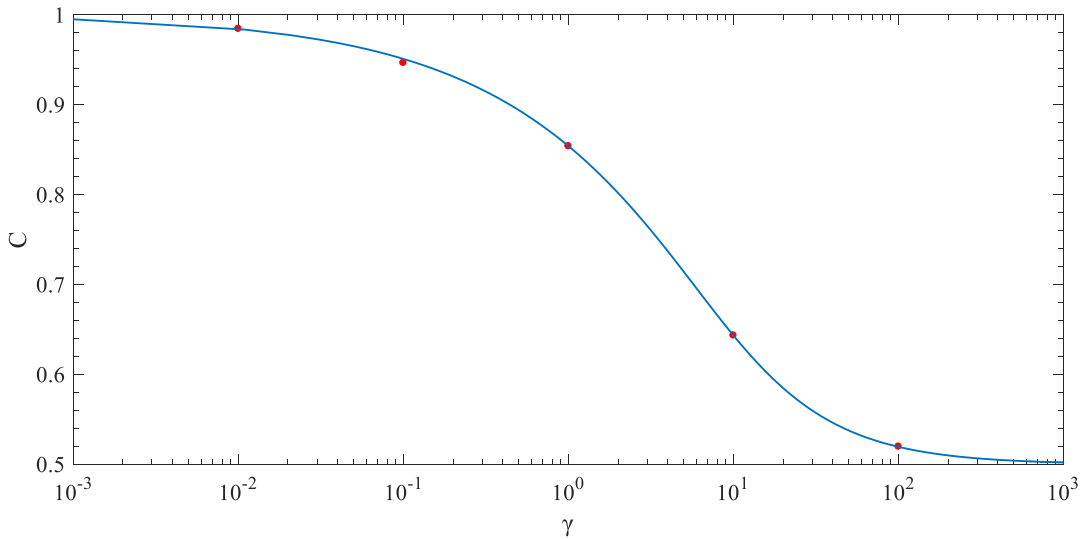}}}
\caption{Exponent ${\cal C}$ $vs$ $\gamma$: 
bosonization prediction (solid line) and numerical results (red dots)  
for $\gamma=0.01,0.1,1,10,100$.} 
\label{fig1}
\end{figure}

Using the known expressions for $\rho^{LD}$ and 
$\rho^{SD}$ (which are $\gamma$-dependent) and matching them with cut-off functions whose parameters have to be optimized, we have been able to set up a very efficient interpolation formula for the 1BDM at any distance $x$ which is given by 
\be
\label{OBRDM_cutoffinterp}
\rho^{INT}(x)= \rho^{SD}(x) \Phi^{SD}(x) + \rho^{LD}(x) \Phi^{LD}(x) \,.
\ee
Having explored several cut-off functions for this optimization, we have finally chosen $\Phi^{SD}$ and $\Phi^{LD}$ 
to be 
\begin{eqnarray}
\Phi^{SD} & = & \left[1-\tanh\left(\frac{\xi}{\alpha}\right) \right] 
\left[1-\tanh\left(\frac{\sqrt{\xi^3}}{\beta}\right)\right] \nonumber \\
\Phi^{LD} &= & \tanh\left(\frac{\xi}{\eta}\right) 
\tanh\left(\frac{\xi}{\omega}\right) \nonumber 
\end{eqnarray} 
where $\xi \equiv \pi n x$ and $\alpha,\beta,\eta$ and $\omega$ 
are coefficients that need to be fixed in terms of $\gamma$, 
but not on $N$. Their best choice comes by minimising the $\chi^2$ in 
a chi-squared test made for small [$\xi<\xi_{min}$]
and large values [$\xi>\xi_{max}$] of $x$,  
where the asymptotic expansions for $\rho(x)$ are known. The test has been done 
for different values of $\xi_{min}$ and $\xi_{max}$, typically 
$\xi_{min} \approx 0.8$ and $\xi_{max} \approx 20$.  
Once parameters $\alpha,\beta,\eta,\omega$ are fixed, we found that the values assumed by $\rho(x)$ in (\ref{OBRDM_cutoffinterp}) 
are in excellent agreement ({\em i.e.} relative percentage errors $< 1\%$) with those obtained for large ($\gamma \geq 100$) and small 
($\gamma \leq 0.01$) couplings at any $x$. Notice the presence of the factor 
$\sqrt{x^{3}}$ in $\Phi^{SD}$ which happens to improve considerably the quality of 
the interpolation for $x \leq 1$.  
Eq. (\ref{OBRDM_cutoffinterp}) is not periodic 
in $x$ with period $L$, 
so we made it symmetric with respect to $L/2$ to better mimic the periodic boundary conditions of the system. Once we fix such a $\rho^{INT}(x)$, we proceed by fixing $N$, computing $\lambda_0$ and check that the result for the given value of $N$ does not depend on the grid in which the interval $[0,L]$ is divided 
\cite{note0}, 
and then we repeat the same procedure for larger values of $N$ at the 
same density. We then extract the exponent ${\cal C}$ from a fit with (\ref{scal}), determining its convergence and error with the given number $N$ (in our case $N=10^3$). Here the possibility of varying $N$ up to large values is important: {\em e.g.,} we get
${\cal C}=0.85391(1)$ for $\gamma=1$. 
In the Tonks-Girardeau case we get ${\cal C}=0.5001(4)$ in agreement with the exact result \cite{LenardTG,ForresterTG}. 
The Tonks-Girardeau result, ${\cal C}=1/2$, confirms the different nature of (non-local) correlation functions of hard-core bosons and ideal fermions, and our outcomes illustrate the crossover from the ideal BEC case to 
the hard-core limit. Let's also mention that 
for $\gamma \sim 10^{-4}-10^{-3}$, realistic for experimentally relevant 
situations, we get ${\cal C} \sim 0.99$, which is very close to $1$. Our results for the behavior of ${\cal C}$ versus $\gamma$ are summarised in Fig.\ref{fig1}.

\section{$1D$ anyons} Let now turn the attention to the case where the system is made of anyons rather than bosons \cite{HaoChen2008}. For a system of $N$ anyons of mass $m$ with contact interactions, the solution $\psi_A$ of the many--body Schr\"odinger equation exhibits a generalised symmetry under the exchange of any pair of particles:
$$\psi_A (\dots , x_{i} , \dots, x_{j} ,\dots) \,=\, e^{i \pi \kappa \, \theta_{ji}} \psi_A (\dots, x_j,\dots,x_i,\dots) \, ,$$
where $\theta_{ji}\equiv sgn(x_j - x_i)$ and $\kappa$ is the so-called \emph{statistical parameter} which runs from $0$ (corresponding to bosons) to $1$ (fermions). The boundary conditions on the wave--functions has to be suitably chosen, because periodic boundary conditions for anyons correspond to twisted boundary conditions for bosons and viceversa \cite{Korepin_anyons}. Hence, imposing twisted boundary conditions and employing the coordinate Bethe ansatz, up to a normalization factor, the eigenfunctions of the system are given by \cite{Batchelor_BA} 
$$\psi_A= e^{-i \frac{\pi \kappa}{2} \sum_{j<k} \theta_{jk}} 
\det[e^{i k_j x_m}]
\prod_{n < l} \left[k_l -k_n - i c' \theta_{ln} \right] \, ,$$
where the indices run from $1$ to $N$ while $c'$ is the renormalized coupling constant given by 
$c'=\frac{c}{\cos(\pi \kappa/2)}$. 
Similarly we set $\gamma' = \frac{\gamma}{\cos(\pi \kappa /2)}$. 
To conveniently obtain the Luttinger parameter $K$ for different values of the coupling constant $\gamma'$ 
and the statistical parameter $\kappa$, we follow the approach in \cite{Korepin_anyons}, where $K$ is given by 
$K = \mathcal{Z}^2$, with $\mathcal{Z} = Z(q)$ where $Z(\lambda)$ is solution of the linear integral equation 
$$Z(\lambda) = 
1+\frac{c'}{\pi} \int_{-q}^{q} \frac{Z(\mu)}{(c')^2 +(\lambda -\mu)^2} \, 
d\mu \, ,
$$  
with $q$ fixed by the Lieb equation for the density of state relative to the anyonic system. 

\section{Hard-core anyons} 
\begin{figure}[t]
\centerline{\scalebox{1.0}{\includegraphics{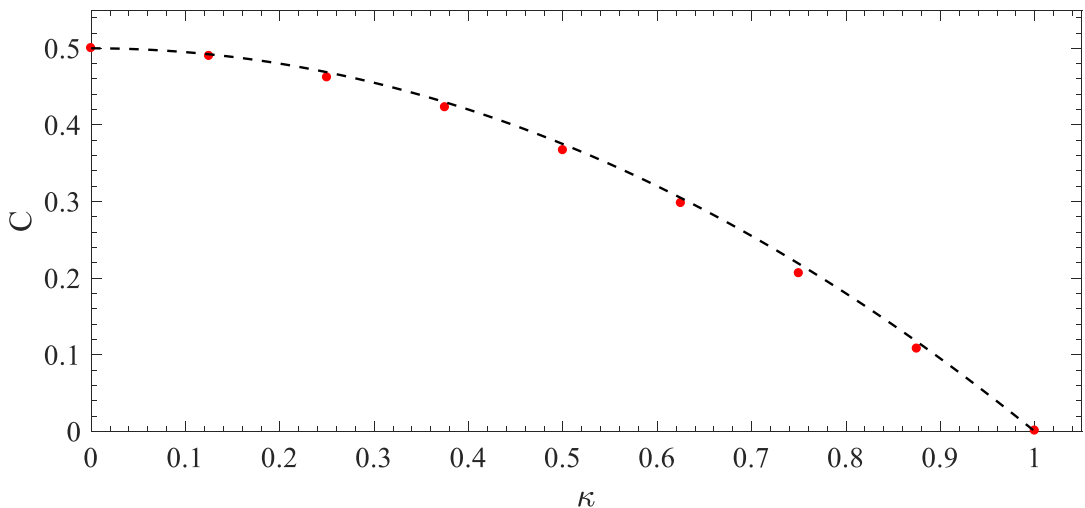}}}
\caption{${\cal C}$ $vs$ $\kappa$. Numerical values from the 
diagonalization of the 1BDM of hard--core anyons (red dots) and 
bosonization result (dashed line): excellent agreement is found.}
\label{fig2}
\end{figure}

With twisted boundary conditions, the $1BDM$ for 
hard-core anyons ($c,\gamma \to \infty$) \cite{Marmorini, Hao2016} is given by 
$\rho^{\kappa}(t)=\det\left[\phi^{\kappa}_{j,l} \right]$ \cite{CalabreseSantachiara}
with $t=2\pi x/L \in [0,2\pi]$, $j,l$ run from $1$ to $N-1$ and
$$\phi^{\kappa}_{j,l}\,=\,\frac{2}{\pi}\int_0^{2\pi} d\tau\, e^{i(j-l)\tau}A(\tau-t) \sin\left(\frac{\tau-t}{2} \right) \sin\left(\frac{\tau}{2}\right)\,\,\,,$$
with $A(\tau - t)=e^{i \pi (1-\kappa)}$ for $\tau < t$ and $A(\tau - t)=1$ for $\tau > t$. 
Proceeding as done before for the LL bosons, we have computed the largest eigenvalue $\lambda_0$ of the 1BDM for different values of the statistical parameter $\kappa$ and number of particles $N$, with $N$ up to $N=241$. The results for ${\cal C}$ are plotted in Fig.\ref{fig2}. 
For $\kappa=0$ (hard-core bosons) one has ${\cal C}=1/2$, while 
for $\kappa=1$ (fermions) one has ${\cal C}=0$, and for all other values the curve monotonically interpolates as expected  
between $1/2$ and $0$ when $\kappa$ increases.

\section{Lieb--Liniger anyons} 
\begin{figure}[t]
\centerline{\scalebox{1.0}{\includegraphics{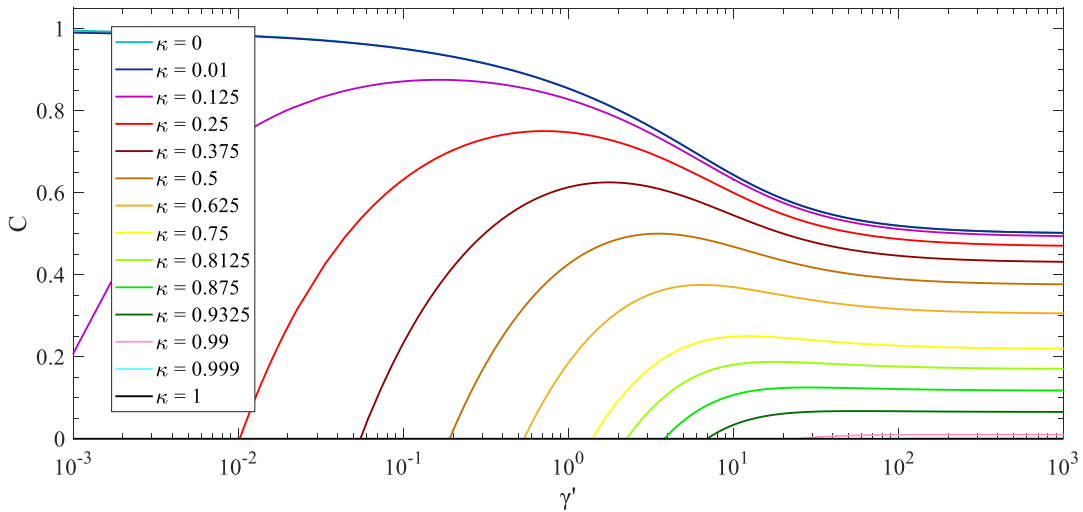}}}
\caption{${\cal C}$ $vs$ $\gamma'$ for varying statistical parameter $\kappa$.}
\label{fig3}
\end{figure}

\begin{figure}[t]
\centerline{\scalebox{1.0}{\includegraphics{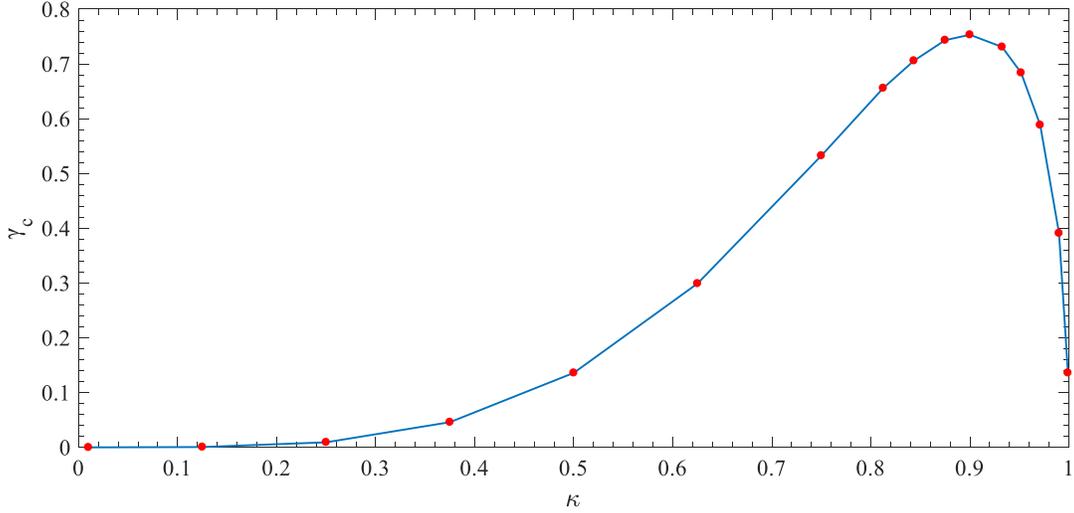}}}
\caption{$\gamma_c$ as a function of $\kappa$.}
\label{fig4}
\end{figure}

The behaviour shown in Fig.\ref{fig2} refers to hard-core anyons. 
For a finite, 
soft-core energy coupling $\gamma$ \cite{CalabreseSantachiara2009} one has the possibility to fully 
interpolate between ${\cal C} = 0$ and $1$: when $\kappa=0$, then 
the LL Bose gas at the coupling constant $\gamma$ is retrieved. 
To study how ${\cal C}$ depends on $\gamma$ and $\kappa$, we resort to the bosonization approach, in light of its successful estimates of ${\cal C}$ both for the LL Bose gas and for hard-core anyons given above. For LL anyons, $\rho(x)$ 
at large distances is given by \cite{Mintchev}
\be
\label{OBRDM_largexAnyons}
\rho(x)=n \, \sum_{m=-\infty}^{\infty} b_m \frac{e^{2i\left(m+\frac{\kappa}{2}\right)k_F x} e^{-\pi i \left(m+\frac{\kappa}{2} \right) {\rm sgn}(x)}}{\left[n L \sin(\pi x/L) \right]^{\left(m+\frac{\kappa}{2}\right)^2 2K+\frac{1}{2K}}} \, ,
\ee
where $b_m$ are non--universal amplitudes. From (\ref{OBRDM_largexAnyons}) one gets for small $k$ and in the thermodynamic limit
\be
\label{mom_distr_anyons}
n(k)\,\propto n \, \sum_{m=-\infty}^{\infty} b_m \frac{(n \pi)^{-\frac{1}{2K}-\left(m+\frac{\kappa}{2} \right)^2 2K}}{\left[k+2k_F \left(m +\frac{\kappa}{2}\right) \right]^{1-\frac{1}{2K}-2K\left(m+\frac{\kappa}{2}\right)^2}} \, .
\ee
As a general consequence of this expression, the momentum distribution in general does not have the maximum at $k=0$: 
there is in fact rather a shift due to the imaginary terms of the 1BDM \cite{Mintchev}. The leading term of $n(k)$ is the one relative to $m=0$ for any $\kappa$ (for $\kappa=1$, also the term $m=-1$ has the same power law behaviour). Hence, we conclude $\lambda_0 \propto N^{1-\frac{1}{2K}-\frac{K\kappa^2}{2}}$, and therefore
the scaling coefficient ${\mathcal C}$ for the LL--anyons is expressed by 
\be
\label{C_powerAnyons}
\mathcal{C}(\kappa,\gamma)\,=\,1-\frac{1}{2K}-\frac{K\kappa^2}{2} \, .
\ee
For $\kappa = 0$ we obtain the result (\ref{res}) for the LL, 
while for $\kappa=1$ (the fermionic limit) 
one gets $\mathcal{C}=0$ since $K=1$ for all $\gamma$, in both cases the correct values. In Fig.\ref{fig3} we plot $\mathcal{C}$ $vs$ $\gamma'$ for different $\kappa$: it is evident that $\mathcal{C}$ is always less than one, as it should be. 
In Fig.\ref{fig3} we do not report of course the negative values of $\mathcal{C}(\kappa)$ because 
there the expression (\ref{mom_distr_anyons}) for the Fourier transform of the 1BDM is not valid 
since the power of $1/x$ in (\ref{OBRDM_largexAnyons}) is greater than $1$. Let's underline that this time ${\cal C}$ is not a monotonic function of $\gamma$. Moreover, it is different from $0$ only for $\gamma$ larger 
than a critical value $\gamma_c$. This result shows once again the singularity related to the bosonic noninteracting limit. A plot of $\gamma_c$ as a function of $\kappa$ is shown in Fig.\ref{fig4}, where it is also 
evident the non-monotonic behavior in $\kappa$, with a maximum around $\kappa \approx 0.9$.

\section{Conclusions} In this paper we have investigated deviations from off-diagonal 
long-range order in a variety of $1D$ systems. For the $1D$ interacting Bose 
gas we have introduced a new interpolating scheme for the one-body density 
matrix based upon the knowledge of large-- and small--distance asymptotical 
behaviours. 
This scheme allows easily to consider systems with large number of particles, 
such as $N=10^3$. 
Our results show that for small interactions the scaling exponent ${\cal C}$ is close to $1$, implying a mesoscopic condensation, {\em i.e.} a value of the ``condensate'' fraction $\lambda_0/N$ appreciable 
at finite values of $N$ (as the ones in experiments with $1D$ ultracold atoms). Finally, we have also shown 
that $1D$ anyons provide the possibility to fully interpolate between ${\cal C}=1$ and $0$ and, moreover, 
that for the behaviour of the exponent ${\cal C}$ for these systems is non-monotonic both with respect to  
the coupling constant and the statistical parameter, revealing the subtleties and pathologies related to the non-interacting limit in $1D$.

\acknowledgments
We thank P. Calabrese and H. Buljan for very useful discussions. Useful 
correspondence with V.E. Korepin is also acknowledged.

\section*{Appendix: Asymptotic expansions} 

At fixed and finite values of $N$ and $L$, in the regime when 
$z \equiv N\gamma \gg 1$, in the translational invariant case, 
one can expand the Bethe eigenfunctions in inverse powers 
of $z$. In this way, in the strong interacting (SI) 
limit, one gets for the 1BDM the asymptotic expansion 
\cite{JimboMiwa,Forrester06}:
\be
\label{OBRDM_strong}
\frac{\rho^{SI}(x)}{n} = \rho_N^{(0)}(x)+\frac{1}{N\gamma} \,\rho_N^{(1)}(x) +
{\cal O}\left(\frac{1}{N\gamma}\right)^2 \, ,
\ee
where $\rho_N^{(0)}$ and $\rho_N^{(1)}$ are expressed in terms of the determinant of certain matrices, see \cite{Forrester06}. 
For $\gamma \rightarrow \infty$ one obtains the formula 
for the 1BDM of a Tonks--Girardeau gas of $N$ particles \cite{LenardTG}
$\rho^{TG}(t)/n=(1/N) \det\left[c_{n,m}(t)\right]_{n,m\,=\,1,\dots,N-1}$, 
where $t = 2\pi x/L \in [0,2\pi]$ and $c_{n,m}=c_{n-m}$ with
$c_i (t) = 2\delta_{i,0}\cos(t)-\delta_{n,1}-\delta_{n,-1}+
(2/\pi) \left[ f(i+1)+f(i-1) + 2i\cos(t/2) \sin(it/2) \right]$
where $f(i)=\frac{\sin(t i/2)}{i}$.
In the weak--coupling limit ({\em i.e.}, if the Luttinger 
parameter $K$ satisfies the inequality $K \gg 1$,  
which amounts to  $\gamma \lesssim 0.1$), 
one can write the 1BDM as \cite{CastinMora}
\be
\label{OBRDM_weak}
\frac{\rho^{WI}(x)}{n} = \exp\left(-\frac{\hbar}{K \sqrt{m \mu}} \int_{0}^{\infty} \, dk \left[1-\cos(k \,x)\right] (v_k)^2 \right) \, ,
\ee
where
$2v_k = [k^2/(k^2 +4\hbar^2/m \mu)]^{1/4} - [(k^2 +4\hbar^2/m \mu)/(k^2)]^{1/4}$ 
and the chemical potential $\mu$ given by 
$\mu = \frac{\hbar^2}{m} n^2 \left[\frac{3}{2} 
e(\gamma) -\frac{\gamma}{2} e'(\gamma) \right]$, with 
$e(\gamma)$ the rescaled ground-state energy 
$e(\gamma)  \equiv \frac{E}{N} \frac{2m}{\hbar^2 n^2}$ obtained solving 
the Lieb integral equations. The expression 
for $\rho^{WI}(x)$ is manifestly non--periodic in $x$: indeed it was originally derived only in the thermodynamic limit where 
$L, N \rightarrow \infty$, as stressed in \cite{CastinMora}. 
To solve this issue we evaluate $\rho^{WI}(x)$ for 
$x \in \left[0 , L/2\right]$, getting all other values of the 1BDM 
for $x \in \left(L/2 , L\right]$ by reflection as 
$\rho(x=L/2 + \delta) = \rho(x=L/2 - \delta)$, 
where $\delta \in (0,L/2]$, so that $\rho(x=L)=\rho(x=0)$. 
This approach turns out to 
be a good way to approximate $\rho(x)$ for studying how $\lambda_0$ 
scales with $N$ when $N$ becomes very large.

For an arbitrary value of the coupling constant $\gamma$, at short distances, 
{\em i.e.} $\mid x \, n \mid\, \ll 1$, in the thermodynamic limit 
the behaviour of the 1BDM is expressed by a Taylor expansion around 
the origin as \cite{Olshanii-short}: 
$\rho^{SD}(x)/n =1 +\sum_{k=1}^{\infty} p_k \mid n\, x \mid^k$,
where the first three Taylor coefficients $p_k$ are given by
$p_1=0$, $p_2=\frac{\gamma e'(\gamma) - e(\gamma)}{2}$ and 
$p_3 \, =\, \frac{\gamma^2}{12} e'(\gamma)$. 
We used only the first three coefficients of this expansion, 
so that the error associated to the truncation is of order $O(x^4)$. 
We have checked the validity of such an approximation for $\rho(x)$ 
by comparing the obtained outcomes 
versus the results for the density matrix at large and small values 
of the coupling ({\em e.g.}, for $\gamma \geq 100$ and 
$\gamma \leq 0.01$) 
from (\ref{OBRDM_strong}) and (\ref{OBRDM_weak}), 
and also versus the results coming from the Tonks--Girardeau expression. 
We found that the relative percentage 
errors 
are well below $1\%$ for 
$\mid x\, n \mid \,\leq 1$.

At large distances, {\em i.e.}  $\mid x\,n \mid \,\gg 1$, 
in the thermodynamic limit the 1BDM can be written as \cite{PanfilCaux-large}
\be
\label{OBRDM_largexTOT}
\frac{\rho^{LD}(x)}{n} = \sum_{m \geq 0} \frac{B_m \, \cos(2 m k_F x)}{(n x)^{2 m^2 K + 1/2K}} \, ,
\ee
where $k_F=\pi n$ is the Fermi--momentum and $B_m$ are 
numerical coefficients which can be determined using the method described in 
\cite{PanfilCaux-large}. One can get already a good approximation 
of the large distance behavior of the 1BDM just by taking the $m=0$. 
Also in this case we have compared the values of the 1BDM 
obtained by (\ref{OBRDM_largexTOT}) versus those relative to weak and 
strong coupling constants and also those coming from the Tonk--Girardeau limit: in all these cases, 
the relative percentage errors remain again always below $1\%$ for 
$\mid x\,n\mid \,\geq 20$.


\begin{thebibliography}{XX}
\bibitem{note1} Since $K=v_F/s$, one has $n(k)\propto k^{\alpha-1}$ 
with $\alpha=ms/2\pi\hbar n$, that, once in the units $\hbar=1$, $m=1/2$ 
as used in \cite{CalabreseABACUS}, 
coincides with Eq.(48) presented there \cite{CalabreseABACUS}. Notice also 
tha putting time $t$ to zero in Eq.(2.6) of Chapter XVIII in \cite{korepin}, 
one gets as well agreement with the results presented in \cite{korepin}.
\bibitem{note0} We checked up to $N=100$ 
that doing the Fourier transform or diagonalizing the 1BDM one gets indeed 
the same results within the numerical precision.
\end{thebibliography}
\end{document}